# Sub-Nyquist Sampling with Optical Pulses for Photonic Blind Source Separation

**Taichu Shi,[1] Yang Qi,[1] Weipeng Zhang,[2] Paul Prucnal,[2] and Ben Wu[1], ***

[1]*Department of Electrical and Computer Engineering, Rowan University, Glassboro, NJ 08028, USA*
2 *Lightwave Communications Laboratory, Department of Electrical Engineering, Princeton University, Princeton, NJ 08544, USA*
*wub@rowan.edu*

**Abstract:** We proposed and demonstrated an optical pulse sampling method for photonic blind source separation. It can separate large bandwidth of mixed signals by small sampling frequency, which can reduce the workload of digital signal processing.

## 1. Introduction

As the wireless communication rapidly growing, the dramatic increase in the use of radio frequency (RF) spectrum will create wireless interference. Blind source separation (BSS) is an effective method to reduce these interferences. It uses a multiple-input, multiple-output (MIMO) receiver to implement [1,2]. Because different outputs contain different information from inputs, through calculating the information only on the receiver, we can separate the signal of interest (SOI) from a set of mixed signals without the aid of information about the source signals or mixing process. Traditional BSS method using digital signal processing (DSP) needs to know the carrier frequency in advance before analog-to-digital conversion (ADC) and BSS can be performed, so it is more effective for lower bandwidth (MHz-GHz). If the bandwidth range exceeds GHz, the real-time blind source separation will bring a considerable burden on the ADC and DSP.

Photonic signal processing provides BSS with a new method that can separate the signals in an analog way without ADC and it can also be implemented on integrated optical circuits [3]. By digitizing only small parts of the signals, and calculating the statistical properties to obtain the corresponding de-mixing matrix, in order to achieve the purpose of separating the SOI from mixed signals. Thus, the sampling frequency can be several orders of magnitude smaller than the bandwidth of the mixed signals, which can greatly reduce the burden on the ADC and DSP. The major challenge is the sampling time. To achieve very short sampling time (pico-second or femto-second) with very low sampling rate (kHz or MHz) for the sampling circuit is unfeasible.

Our research generates a series of optical pulses and modulate the RF signal by using these pulses, utilizes the under-sampling properties of photonic BSS for wideband RF signal processing. By calculating the statistical information of a small part of the mixed signal after optical sampling, we can obtain a de-mixing matrix that is sufficient to accurately separate the SOI. The optical pulse width is in the order of 100 femto-seconds and the ratio of the optical sampling signal for calculating the statistical properties to the mixed signal is from $1/10^2$ to $1/10^5$.

## 2. Principle and Experimental Setup

Fig. 1 (a) shows the schematic diagram of photonic BSS system. The photonic BSS is implemented by a MIMO receiver. The mixed signals are represented by [4]:

$$\boldsymbol{X} = \boldsymbol{AS} \text{ or, } \begin{bmatrix} x_1 \\ x_2 \end{bmatrix} = \begin{bmatrix} a_{11} & a_{12} \\ a_{21} & a_{22} \end{bmatrix} \begin{bmatrix} s_{soi} \\ s_{int} \end{bmatrix} \quad (1)$$

In this formula, $\boldsymbol{A}$ represents the mixing matrix. $x_1$ and $x_2$ are received mixed signals from MIMO antennas. The mixing matrix describes how SOI and interference are mixed. Here in our experiment, our SOI uses a 16QAM signal and a Gaussian noise is used as interference. The black dots in Fig. 1 (b) represent the real parts of received signals $x_1$ and $x_2$. The four straight lines indicate that it is a 16QAM signal, and the dots on each line are distributed in Gaussian. To separate the SOI from interference, we need to find $\boldsymbol{A^{-1}}$ through principal component analysis (PCA) and independent component analysis (ICA):

$$\boldsymbol{A^{-1}} = \boldsymbol{V U \Sigma U^{-1}} \quad (2)$$

Where $\boldsymbol{U \Sigma U^{-1}}$ represents the PCA, and $\boldsymbol{V}$ represents the ICA. The mixed signal can be regarded as the received signal with different weights. These weights can be achieved by adjusting the attenuation of the photonic circuit.

$$x_{PCA} = \cos(\theta) x_1 + \sin(\theta) x_2 \quad (3)$$

Thus, the second order moment of mixed signal can be represented by a function of $\theta$ (shows in Fig. 1 (b)):

$$E(x_{PCA}^2) = q_1 + q_2 \cos[2(\theta - \theta_0)] \quad (4)$$

In Equation 4, $E(x_{PCA}^2)$ is the time average of $x_{PCA}^2$. $\theta_0$ shows the direction of first principal component. Therefore, we can calculate the magnitude of the first and second principal component. They are $q_1 + q_2$ and $q_1 - q_2$. We can then use different values to determine the three unknown parameters, $q_1, q_2$ and $\theta_0$. The PCA procedure then can normalize and whiten the mixed signal by calculating $X' = [x'_1, x'_2] = U\Sigma U^{-1}X$, where:

$$U = \begin{bmatrix} \cos(\theta_0) & -\sin(\theta_0) \\ \sin(\theta_0) & \cos(\theta_0) \end{bmatrix} \quad \text{and} \quad \Sigma = \begin{bmatrix} 1 & 0 \\ 0 & \sqrt{\frac{q_1+q_2}{q_1-q_2}} \end{bmatrix} \quad (5)$$

Similarly, we can use 4$^{th}$ order moments (kurtosis) of the whitened signals to perform ICA:

$$E(x_{ICA}^4) = p_1 + p_2 \cos[2(\phi - \phi_0)] + p_3 \cos[4(\phi - \phi_0)] \quad (6)$$

$$V = \begin{bmatrix} \cos(\phi_0) & -\sin(\phi_0) \\ \sin(\phi_0) & \cos(\phi_0) \end{bmatrix} \quad (7)$$

After all the matrix are calculated, we can get the de-mixing matrix and separate SOI from interference.

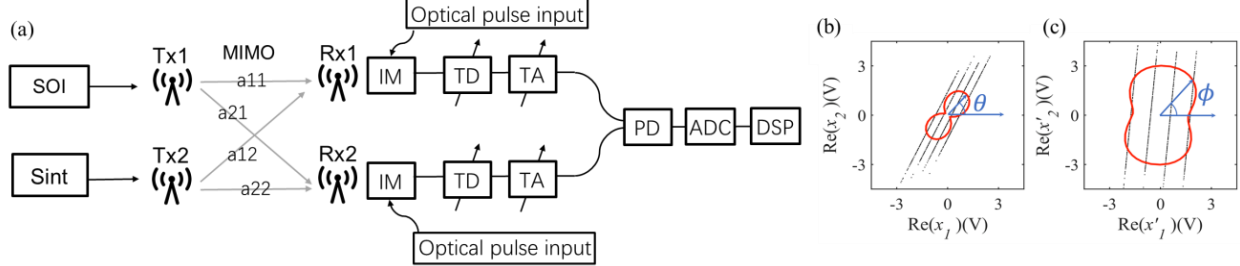

Fig. 1. (a)BSS system schematic (IM: optical intensity modulator, TD: tunable optical delay, TA: tunable optical attenuator, PD: photodetector, ADC: analog to digital conversion, DSP: digital signal processing) (b) black dots: mixed signals; red line: 2$^{nd}$ order moment (c) black dots: whitened signals; red line: 4$^{th}$ order moment.

## 3. Results and Analysis

We generate optical pulses by using optical intensity modulator. They can also be generated by a mode-locked laser. As shown in Fig. 2 (a), the pulse width is 5 ns. The SOI we use here is a random binary signal with 200Mbps data rate (shown in Fig. 2 (b)) and the interference is a Gaussian noise. Fig. 2 (c) is the sampled signal by optical pulse. The eyediagram of the sampled signal is still very clear. Fig. 2 (d) shows the theoretically calculated 2$^{nd}$ order moment and the mixed signals used in the experiment. Fig. 2 (e) shows the comparison of theoretical calculated 2$^{nd}$ order moment and the experimental 2$^{nd}$ order moment which is obtained by calculating the optical samples. The black points are different values of $\theta$ we chose to calculate the de-mixing matrix. When $\theta = 0°$, it represents $x_1$ and 90° for $x_2$.

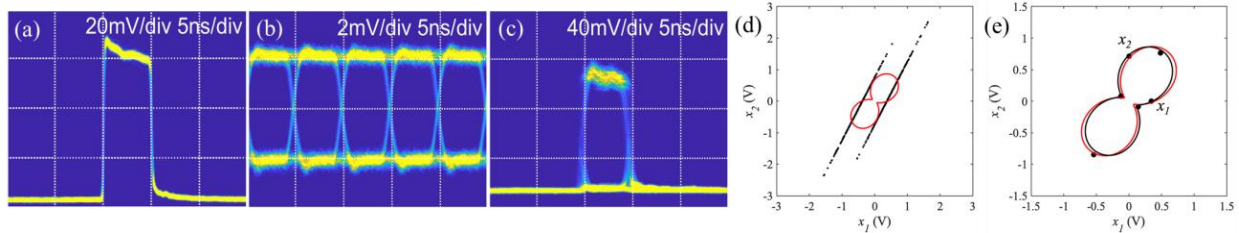

Fig. 2. (a) Optical pulses (b) Random binary signal of 200Mbps (c) Sampled Signal (d) Mixed signal and 2$^{nd}$ order moment (e) Comparison of theoretical 2$^{nd}$ order moment (black) and experimental 2$^{nd}$ order moment (red).

## 4. Conclusion

We proposed and experimentally demonstrated a new optical sampling method for photonic blind source separation. It generates a series of optical pulses and use them to pre-sample the mixed signal before ADC, which can achieve separating the SOI form interference by using low sampling rate for high data rate signals. It reduces the workload of ADC and DSP by orders of magnitude.